\documentclass[
reprint,
superscriptaddress,
 amsmath,amssymb,
 aps,
]{revtex4-1}

\usepackage{graphicx}
\usepackage{dcolumn}
\usepackage{bm}
\usepackage{color}
\usepackage{MnSymbol}

\definecolor{mygreen}{rgb}{0,0.85,0.45}
\definecolor{myred}{rgb}{1.0,0.44,0.41}
\definecolor{mypurple}{rgb}{0.57,0.49,0.84}
\definecolor{myblue}{rgb}{0,0.53,0.77}

\begin{document}


\title{Autothermotaxis of volatile drops}
 
\author{Pallav Kant}
\thanks{These authors contributed equally}
\affiliation{Physics of Fluids Group, Max Planck Center Twente for Complex Fluid Dynamics, University of Twente, 7500 AE Enschede, The Netherlands}
\affiliation{Bullard Laboratories, Department of Earth Science, University of Cambridge}

\author{Mathieu Souzy}
\thanks{These authors contributed equally}
\affiliation{Physics of Fluids Group, Max Planck Center Twente for Complex Fluid Dynamics, University of Twente, 7500 AE Enschede, The Netherlands}
\affiliation{INRAE, Aix Marseille Univ, RECOVER, Aix-en-Provence, France}

\author{Nayoung Kim}
\thanks{These authors contributed equally}
\affiliation{Physics of Fluids Group, Max Planck Center Twente for Complex Fluid Dynamics, University of Twente, 7500 AE Enschede, The Netherlands}

\author{Devaraj van der Meer}
\affiliation{Physics of Fluids Group, Max Planck Center Twente for Complex Fluid Dynamics, University of Twente, 7500 AE Enschede, The Netherlands}

\author{Detlef Lohse}
\affiliation{Physics of Fluids Group, Max Planck Center Twente for Complex Fluid Dynamics, University of Twente, 7500 AE Enschede, The Netherlands}
\affiliation{Max Planck Institute for Dynamics and Self-Organization, Am Fassberg 17, 37077 G{\"o}ttingen, Germany}

\date{\today}

\begin{abstract}
When a drop of a volatile liquid is deposited on a uniformly heated wettable, thermally conducting substrate, one expects to see it spread into a thin film and evaporate. Contrary to this intuition, due to thermal Marangoni contraction the deposited drop contracts into a spherical-cap-shaped puddle, with a finite apparent contact angle. Strikingly, this contracted droplet, above a threshold temperature, well below the boiling point of the liquid, starts to spontaneously move on the substrate in an apparently erratic way. We describe and quantify this self-propulsion of the volatile drop. It arises due to spontaneous symmetry breaking of thermal-Marangoni convection, which is induced by the non-uniform evaporation of the droplet. Using infra-red imaging, we reveal the characteristic interfacial flow patterns associated with the Marangoni convection in the evaporating drop. A scaling relation describes  the correlation between  the moving velocity of the drop and the apparent contact angle,  both of which increase with the substrate temperature. 
\end{abstract}
\maketitle

Inkjet printing \cite{lohse2022fundamental, calvert2001inkjet, shimoda2003inkjet, park2006control, derby2010inkjet}, spray cooling \cite{grissom1981liquid, kim2007spray}, self-assembly techniques \cite{jiang1999single, rabani2003drying, narayanan2004dynamical, schnall2006self, marin2012building} and forensic science \cite{smith2020new, attinger2013fluid} 
all rely on controlled evaporation of a droplet on a surface. In their seminal work, Deegan {\it et al.} \cite{deegan1997capillary, deegan2000contact, deegan2000pattern} have shown  that non-uniform evaporation of a droplet deposited on a surface establishes a capillary flow that continuously compensates for enhanced evaporative losses at the edge of the droplet. The corresponding flow-field, however, can be affected by interfacial (Marangoni) flow induced by a gradient in temperature along the interface, also due to the non-uniform evaporation \cite{lohse2022fundamental, lohse2020nat,wang2022, gelderblom2022}. Interestingly, the direction of thermal gradient and the consequent Marangoni flow is critically influenced by the heat exchange between the droplet and the substrate \cite{ristenpart2007influence}.

In this paper we demonstrate that spontaneous symmetry breaking of this Marangoni flow can lead to self-propulsion of a volatile droplet on a warm, thermally conducting substrate. Note that unlike the motion of a droplet in the presence of an extrinsic gradient such as thermal \cite{brochard1989motions, brzoska1993motions, yarin2002motion}, chemical \cite{chaudhury1992make, dos1995free, daniel2001fast}, electrical \cite{gunji2005self} or photo-chemical \cite{ichimura2000light}, the spontaneous motion of the volatile droplets shown here emerges due to intrinsically unstable thermal Marangoni convection. Accordingly, parallels can be drawn between our system and the self-propulsion of dissolving droplets (autochemotaxis), which generate propulsion forces due to intrinsically unstable solutal-Marangoni flows in an isotropic surrounding medium \cite{michelin2013spontaneous,izri2014self,jin2017chemotaxis,michelin2023self,maass2016,hokmabad2021}. Therefore we call the phenomenon {\it autothermotaxis}.

\begin{figure}[!]
\centering
\includegraphics[clip, trim=0cm 0cm 0cm 0cm, width=0.45\textwidth]{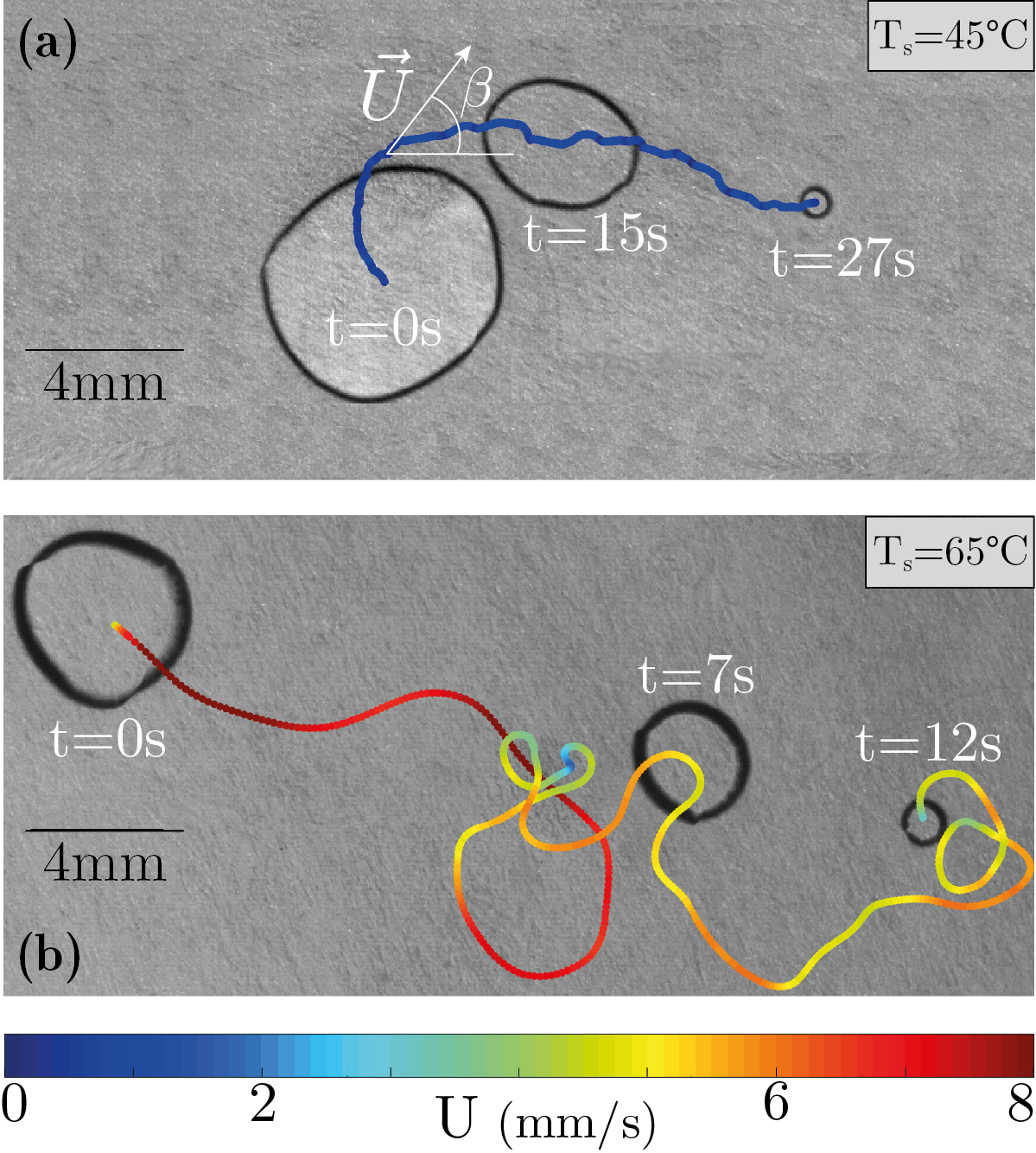}
\caption{Chronophotography of volatile ethanol droplets of volume 4 $\mu$l, deposited on a horizontal heated sapphire plate at temperature $T_\mathrm{s}$ = (a) 45$^\circ$C and (b) 60$^\circ$C. Following a short transient period after deposition, the droplet self-propels (autothermotaxis) until complete evaporation. Trajectories are overlaid on the images and color-coded by instantaneous velocity $U$. Droplets move faster and have an increased tendency to reorient on warmer substrates.}
\label{fig:fig1}
\end{figure}

\begin{figure}
\centering
\includegraphics[clip, trim=0cm 0cm 0cm 0cm, width=0.45\textwidth]{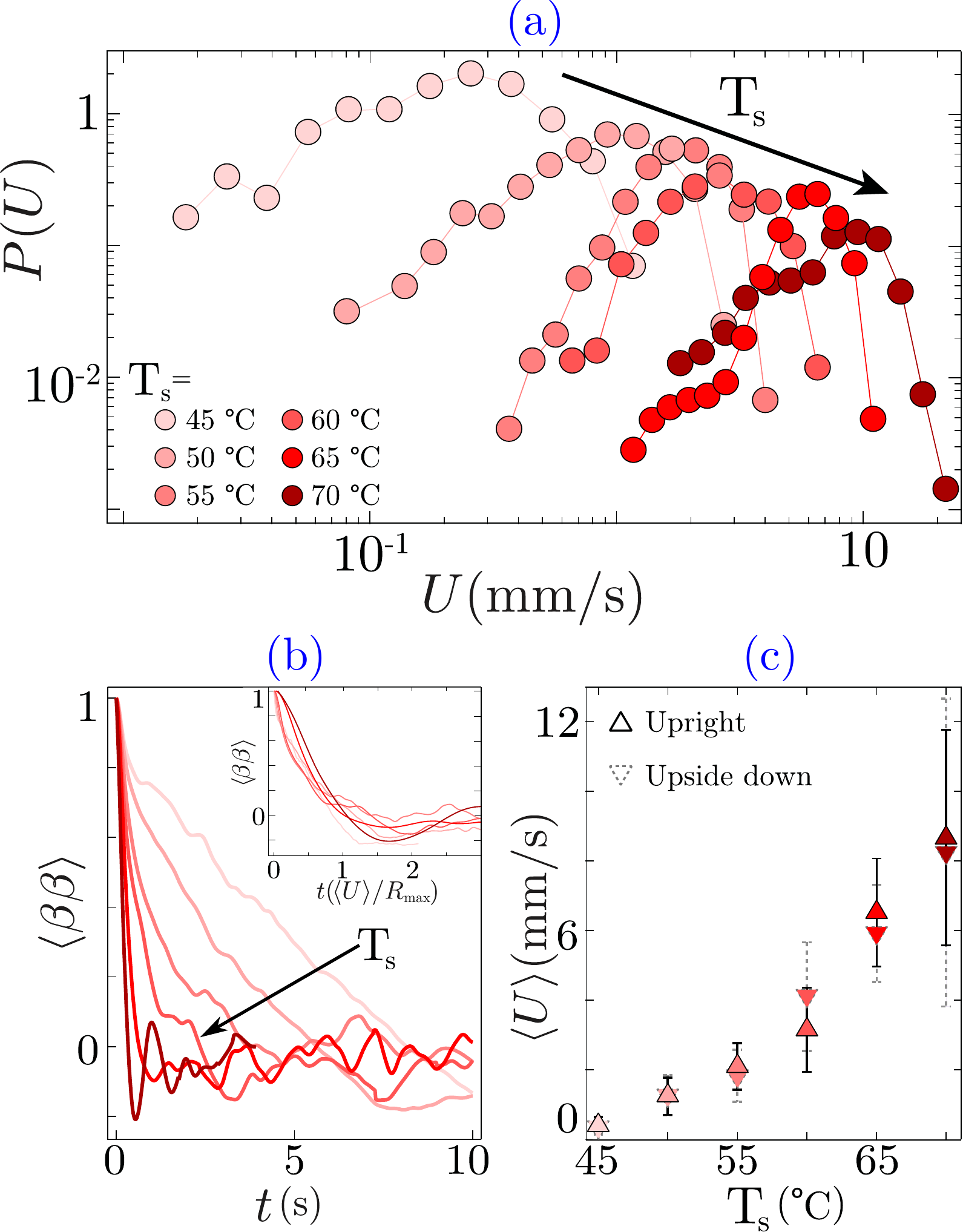}
\caption{(a) Probability distribution function of velocity $P(U)$ of ethanol droplets on a warm sapphire plate for different substrate temperatures $T_\mathrm{s}$. (b) Autocorrelation function of the orientation angle $\beta$ with increasing $T_\mathrm{s}$. Inset: Data collapse when non-dimensionalizing time with the maximum droplet radius $R_\mathrm{max}$ and the mean velocity $\langle U \rangle$. (c) Comparison of the mean velocity $\langle U \rangle$ of the droplet measured in upright and upside-down configurations. Error bars represent the standard deviation of the velocities. Close match between the two configurations highlights the insignificant role of gravity in the droplet motion.}
\label{fig:fig2}
\end{figure}

\begin{figure}
\centering
\includegraphics[clip, trim=0cm 0cm 0cm 0cm, width=0.45\textwidth]{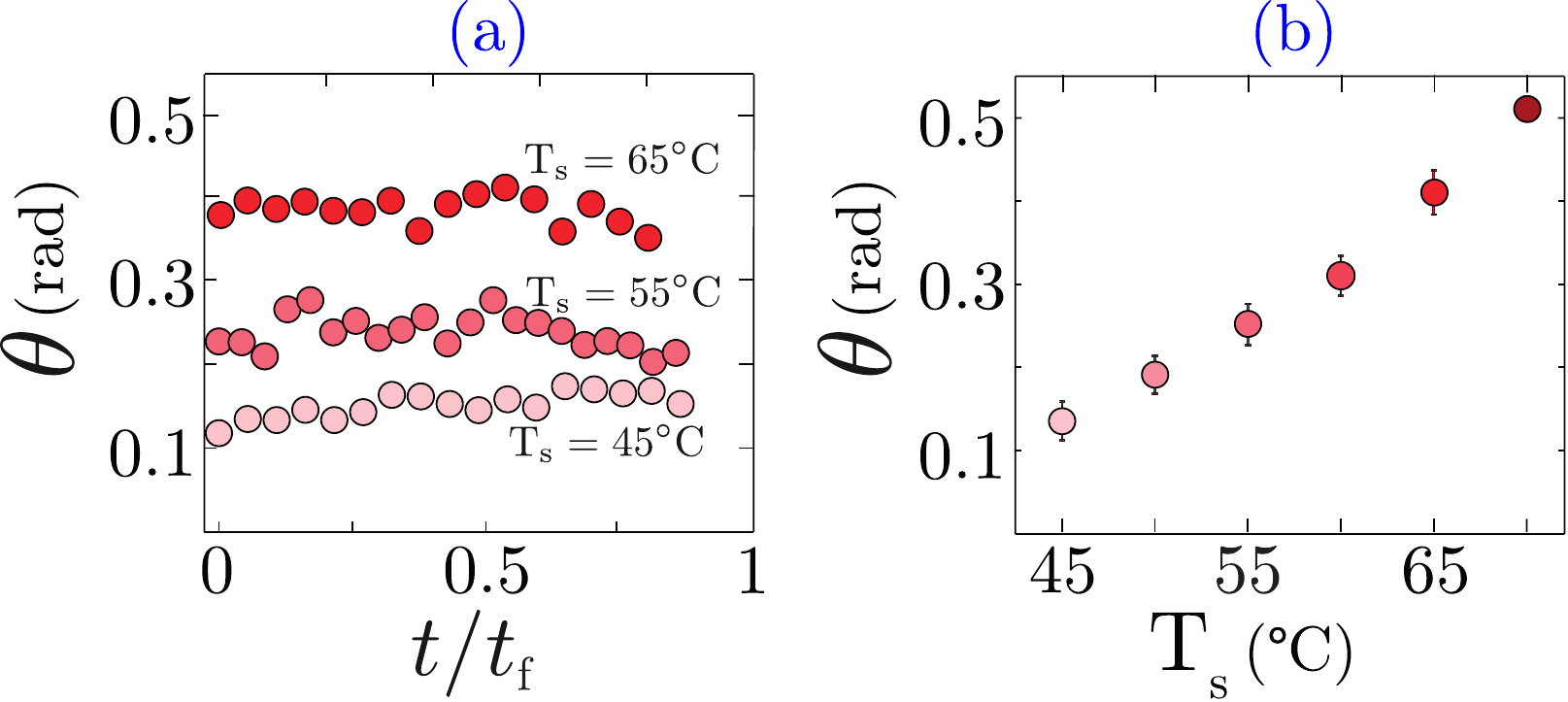}
\caption{(a) Temporal measurement of apparent contact angle $\theta$ during the lifetime of the droplet for different substrate temperature $T_\mathrm{s}$. Time is non-dimensionalized using the droplet lifetime $t_f$. (b) Increase in the contact angle $\theta$  of a self-propelling ethanol droplet with $T_\mathrm{s}$.}
\label{fig:fig3}
\end{figure}

Deposition of a small volume of pure ethanol ($\sim 4\,\mu$l) on a horizontal sapphire plate at room temperature yields a thin film with a vanishing contact angle $\theta \sim 0^\circ$, which evaporates slowly over time. This behaviour is expected for  a substrate-liquid pair for which the spreading parameter $S = \gamma_{\rm{sv}} - \left(\gamma_{\rm{sl}} + \gamma_{\rm{lv}} \right)$ is positive; here $\gamma_{\rm{ij}}$ denotes the interfacial tension between \emph{i}-\emph{j} phases, namely, substrate, liquid and vapor. However, despite the favourable wetting conditions, when the same volume of ethanol is deposited on a uniformly heated sapphire plate it contracts into a spherical-cap-shaped droplet, exhibiting a finite apparent contact angle. For a droplet deposited on a substrate such that their thermal conductivity ratio $k_\mathrm{R} = k_\mathrm{sub}/k_\mathrm{drop} \gg 2$ (here for ethanol on sapphire $k_\mathrm{R} = 228$), this contraction is caused by the evaporation-induced thermal Marangoni flow that is directed radially inwards along the droplet interface \cite{ristenpart2007influence, tsoumpas2015effect, shiri2021thermal}. The resulting apparent droplet contact angle after the thermal Marangoni contraction is shown in Fig.\,\ref{fig:fig3}; for fixed surface temperature it is roughly constant in time, but becomes larger with increasing surface temperature, reflecting the stronger thermal Marangoni flow.

We now come to the main finding of this paper, namely that beyond a threshold temperature $T_\mathrm{onset} \sim 45^\circ$C, the Marangoni contracted ethanol droplets spontaneously start to move in a seemingly erratic  fashion along the substrate. This self-propulsion of an ethanol droplet persists until it has evaporated entirely at time $t = t_\mathrm{f}$; see Fig.\,1 and Supplementary Video. We emphasize that the self-propulsion of a volatile droplet as reported here occurs at temperatures much lower than its boiling point $T_\mathrm{b}$, which  for ethanol is $T_\mathrm{b} = 78^\circ$C. Therefore, it bears no commonality with mobile Leidenfrost droplets on strongly superheated surfaces \cite{linke2006self, quere2013, bouillant2018leidenfrost}. Additionally, the observed self-propulsion of droplets is not exclusive to the ethanol-sapphire combination; it is also evident in various combinations of volatile liquids and substrates; see Table \ref{tab:table}. However, to keep the discussion concise, we will focus on the fundamental characteristics of this phenomenon using the ethanol-sapphire system.

\begin{table}[htb]
  \begin{center}
	\centerline{\includegraphics[width=0.48\textwidth]{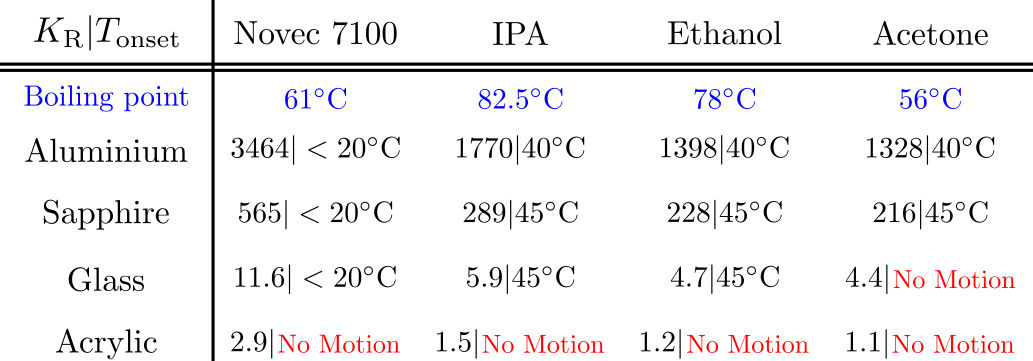}}
\caption{Thermal conductivity ratio $k_\mathrm{R} = k_\mathrm{sub}/k_\mathrm{drop}$ and threshold temperature $T_\mathrm{onset}$ for various liquid-substrate combinations, revealing the generic nature of autothermotaxis.}
\label{tab:table}
\end{center}
\end{table}

\begin{figure*}[htb]
\centering
\includegraphics[clip, trim=0cm 0cm 0cm 0cm, width=0.8\textwidth]{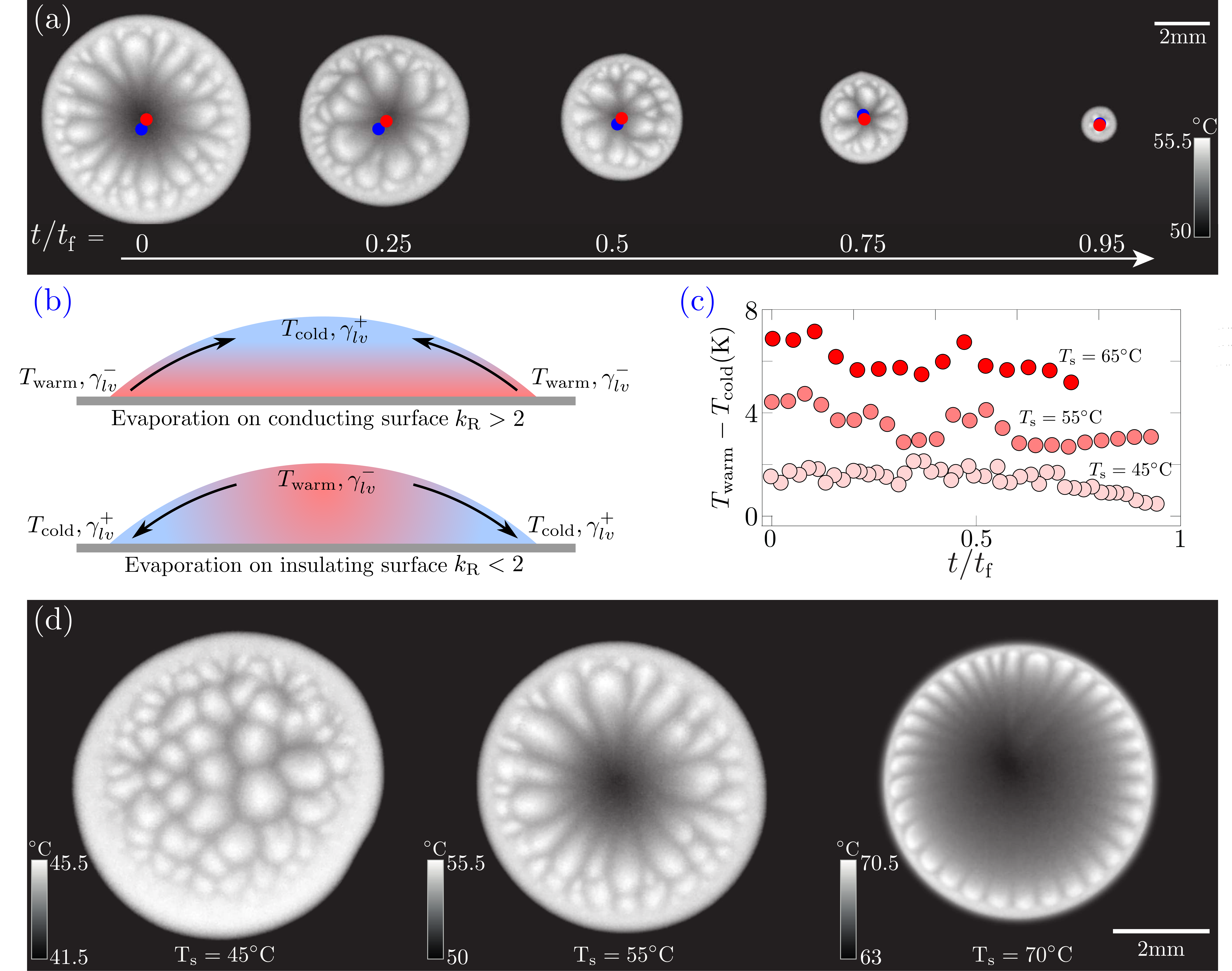}\\
\caption{Infra-red imaging of self-propelling droplets: (a) Interfacial thermal map (top-view) of a moving ethanol droplet while it evaporates on a sapphire substrate at $T_\mathrm{s} = 55^\circ$C. The bright region close to the contact line is the warmest part of the droplet, whereas the dark region near the apex is the coldest. Red and blue dots corresponding respectively to the centroid of droplet footprint and convection cells reveal the spontaneous symmetry breaking of the Marangoni flow. 
(b) Sketches of the thermal gradient and the resulting Marangoni flows for an evaporating droplet on a heated conducting ($k_\mathrm{R} > 2$) and on an  insulating ($k_\mathrm{R} < 2$) substrate. (c) Temperature difference between the warmest and coldest part of a self-propelling ethanol droplet for various $T_\mathrm{s}$. (d) Variation in interfacial temperature patterns with substrate temperature $\rm{T_s}=45,\,55,\,70^\circ$C.}
\label{fig:fig4}
\end{figure*}

Overlaid images in Fig.\,1 illustrate the contrasting behaviours of a self-propelling ethanol droplet at two different substrate temperatures. At lower temperatures, the self-propelling droplet moves slowly and uniformly (Fig.\,1a), whereas fast and chaotic movements are observed at higher temperatures (Fig.\,1b). The distribution of propulsion velocities $U$ measured at different substrate temperatures, shown in Fig.\,2a, demonstrates this systematic shift towards faster velocities with increasing substrate temperature. Note that at elevated temperatures, the motion of a droplet is largely unaffected by surface heterogeneities (roughness), whereas  self-propelling droplets at low substrate temperatures are more prone to pinning.
Further, as highlighted in Fig.\,\ref{fig:fig2}b, on warmer substrates the fast motion of a droplet is also coupled with an increased tendency for abrupt re-orientations. This type of motion is reminiscent of a `run-and-tumble' movement observed in self-phoretic biological systems and artificial microswimmers \cite{berg2018random, polin2009chlamydomonas, hokmabad2021}. We quantify the tendency of a moving droplet to re-orient itself by measuring the auto-correlation function $\langle \beta \beta \rangle = \frac{\langle\beta(t' + t) \beta(t)\rangle_{t'}}{\langle\beta(t')^2\rangle} $ of its orientation angle $\beta$; see Fig.\,1. As shown in Fig.\,2b, the quick decay of the autocorrelation function at higher temperatures demonstrates that faster-moving droplets lose their directionality more rapidly. As shown in the inset of Fig.\,2b, by non-dimensionalizing time, using mean velocity $\langle U \rangle$ and maximum droplet radius $R_\mathrm{max}$, a collapse of $\langle \beta\beta \rangle$ indicates that irrespective of the substrate temperature a moving droplet re-orients itself after moving typically about one radius. Importantly, since a reorientation event consists of droplet slowing down while changing direction and then re-accelerating along a new path, the droplet velocity on warmer substrates varies dramatically over its lifetime. As a result, both the mean and standard deviation of the velocity of a droplet increase with temperature. A close comparison between the mean velocity $\left< U \right>$ of moving droplets in upright and upside-down orientations (Fig.\,2c) shows that effects arising from density changes due to temperature variations (i.e., thermal convection) within the droplet play no significant role in the overall dynamics, reflecting that the corresponding Archimedes number (the ratio between buoyancy forces due to temperature-induced density differences and viscous forces) 
\cite{li2019prl-yaxing} is smaller than 1, namely Ar $= g h^3 \rho_\mathrm{warm}(\rho_\mathrm{cold} - \rho_\mathrm{warm})/\mu^2$, see Supplementary Material.


The two main arising questions are:  Why does the volatile droplet move spontaneously on the  uniform surface, without any extrinsic asymmetry? And how (it at all) is this spontaneous motion of the evaporating droplet related to  its thermal Marangoni contraction? 

To reveal the underlying physical mechanism, we make use of infra-red (IR) imaging, allowing quantitative measurement of the interfacial thermal activity. Fig.\,\ref{fig:fig4}a shows the interfacial thermal map of a self-propelling ethanol droplet deposited on a sapphire plate at $T_\mathrm{s} = 55^\circ\mathrm{C}$ at different instances of its lifetime.
As anticipated, for a droplet substrate pair with $k_\mathrm{R} \gg 2$, the conduction between the substrate and the liquid suppresses the evaporative cooling effect near the contact line, thus it is the warmest region $\left(T_\mathrm{warm} \right)$ on the interface. In contrast, the apex of the droplet is the coldest region $\left(T_\mathrm{cold} \right)$. Our measurements indicate that the difference in the temperature of these two regions on the droplet interface $\Delta T = T_\mathrm{warm} - T_\mathrm{cold}$ remains nearly constant throughout the droplet's lifetime while increasing with a rise in the substrate temperature; see Fig.\,\ref{fig:fig4}c. Correspondingly, the persistent stronger Marangoni flow along the interface facilitates the enhanced contraction of a droplet on warmer substrates.

Strikingly, these IR images also reveal the presence of non-stationary convective patterns on the droplet interface; see also Supplementary movies. The emergence of such patterns is a signature of a flow-instability of the evaporating droplet that is driven by thermal Marangoni flow originating from the variations of surface tension with temperature \cite{davis1987thermocapillary, schatz2001experiments, sefiane2008self, karapetsas2012convective, shi2017marangoni}. 
Importantly, the resultant unsteady patterns are also indicative of an unsteady flow field inside the droplet, exhibiting spontaneous symmetry breaking. This spontaneous symmetry breaking of the flow field manifests itself through the asymmetric arrangement of convective cells. This is confirmed by comparing the centroid of the interfacial convection cells with the centre of the droplet footprint, as shown in Fig.\,\ref{fig:fig4}a. The spontaneous symmetry breaking of the unsteady flow field provides the impetus for the self-propulsion of a droplet, similarly as for the self-propulsion of a dissolving droplet by solutal Marangoni flow \cite{michelin2013spontaneous, izri2014self, jin2017chemotaxis, michelin2023self, maass2016, hokmabad2021}. 

We emphasize that there is no causal relationship between the Marangoni contraction of droplets and the Marangoni instability of evaporating droplets. These two distinct mechanisms govern two different aspects of the thermal Marangoni flow, namely, contraction and lateral motion by spontaneous symmetry breaking.


\begin{figure}
\centering
\includegraphics[clip, trim=0cm 0cm 0cm 0cm, width=0.38\textwidth]{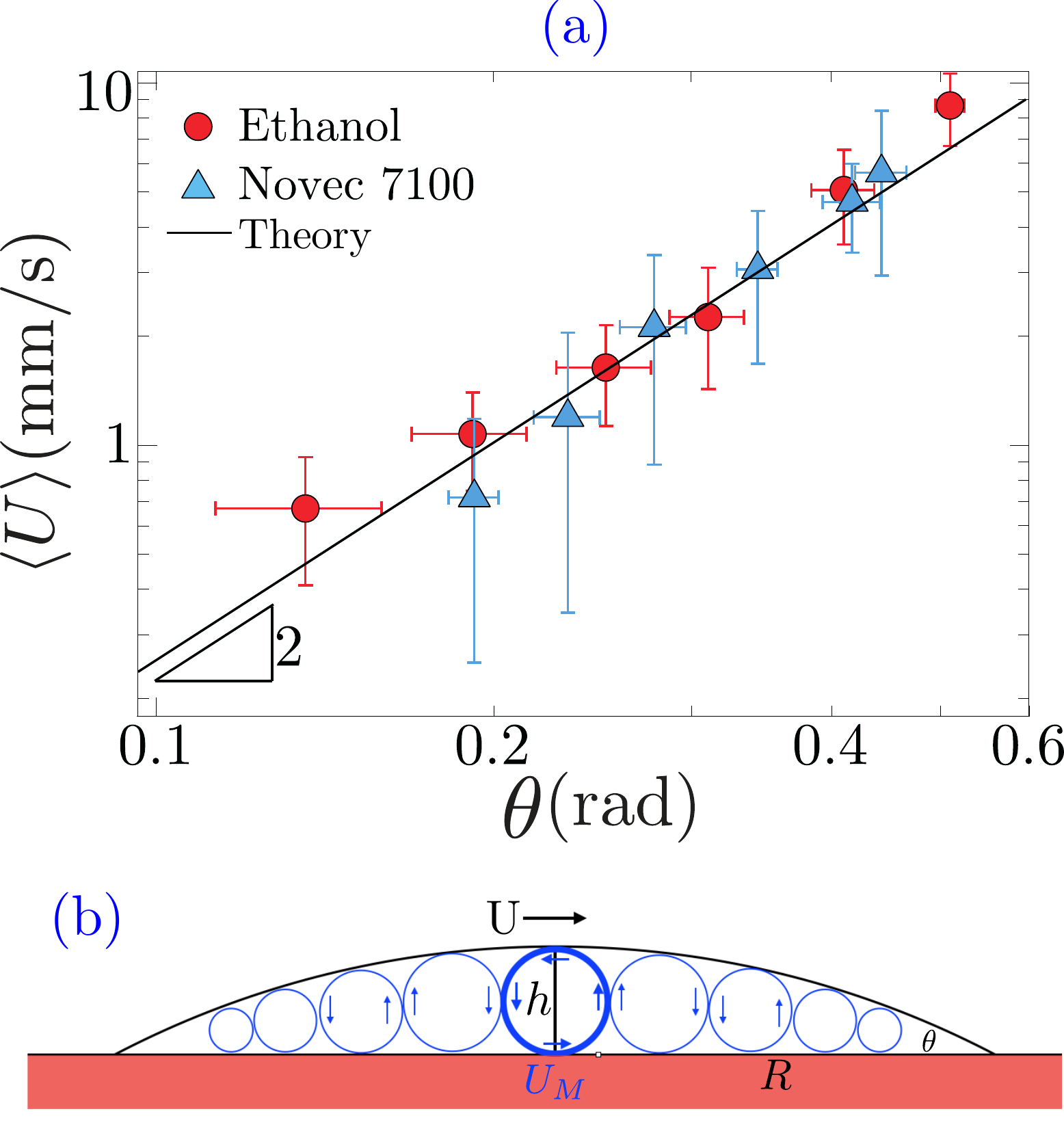}\\
\caption{(a) Mean droplet velocity  $\langle U\rangle$ measured as a function of the apparent contact angle $\theta$ for ethanol and Novec 7100 droplets deposited on a warm sapphire plate. The solid black-line is the expected scaling $\langle U \rangle \sim \theta^2$, cf.\ eq.\ (\ref{correlation}). Error bars represent the standard deviation over multiple runs. (b) Sketch of the Marangoni rolls inside the moving drop, after spontaneous symmetry breaking. 
See main text for explanations.}
\label{fig:fig5}
\end{figure}

The Marangoni instability in the evaporating ethanol droplet manifests itself through distinct convective patterns, depending on the substrate temperature. Fig.\,\ref{fig:fig4}d shows a comparison of distinct patterns exhibited by ethanol droplets at different substrate temperatures. At lower substrate temperatures, the convective pattern includes irregularly shaped convection cells distributed randomly over the droplet interface, which is also reminiscent of disordered Marangoni-convection observed for a very thin liquid layer heated from the bottom \cite{cerisier1996topological}. In contrast, at higher temperatures,  flower-like patterns with azimuthally arranged convection cells appear. The aspect ratio and thus the apparent contact angle $\theta \sim h/R$, where $h$ is the drop height and $R$ its contact radius, determines the morphology of the convective pattern, as also demonstrated previously for heated liquid films in circular containers \cite{koschmieder1990surface} or for Rayleigh-B\'enard convection \cite{wang2020prl}. We also performed experiments with droplets with pinned contact lines to confirm the role of droplet aspect ratio in selecting convective patterns. In such a configuration, the droplet aspect ratio decreases steadily as the drop evaporates. Accordingly, this induces a striking transition of convective patterns; see Supplementary Material.

Finally, we provide the correlation between the propulsion velocity and the apparent droplet contact angle. 
We obtain this by balancing the driving force of the droplet movement with the overall friction force, see Fig.\ \ref{fig:fig5}b:
The Marangoni driving occurs after spontaneous symmetry breaking; thereafter one Marangoni convection roll  (or a few in 3D; see the bold blue roll in Fig.\ \ref{fig:fig5}b) dominates the others, whose net-forces on the droplet cancel out. As the aspect ratio of the convection rolls is roughly unity \cite{wang2020prl} the area $A_M$ of the dominating Marangoni roll is determined by the Marangoni roll height $h$, i.e., $A_M \sim h^2$ and the resulting frictional Marangoni  force on the droplet thus is $F_M \sim A_M U_M \mu /h$, with $\mu$ being the dynamic viscosity and $U_M \sim { \Delta \sigma \over \mu}  \sim {1\over \mu } {d\sigma \over dT} \Delta T $ being the Marangoni velocity. 
In contrast, the overall viscous friction force on the droplet with its overall velocity $U$ acts on the whole droplet area $A_D \sim R^2$, i.e., $F_D\sim A_D \mu U/h$. The force balance $F_M \sim F_D$ thus results in $A_M U_M \sim A_D U$ or in 
the desired correlation 
 \begin{equation}
\left< U \right> \sim U \sim U_M { h^2\over R^2 }
\sim {1\over \mu } {d\sigma \over dT} \Delta T \  \theta^2,
\label{correlation}
\end{equation}
 where we have assumed that $h/R = \tan \theta \approx \theta$ is small. 
  Eq.\ (\ref{correlation}) thus indeed reflects the observed quadratic dependence of the average velocity $\left<U\right>$ on the apparent contact angle $\theta$ of the Marangoni contracted droplet as observed in Fig.\  \ref{fig:fig5}a.
Here, we have ignored the slight dependence of $\Delta T$ on the surface temperature (Fig.\ \ref{fig:fig4}c). 
Incorporating this dependence $\Delta T(T_s)$ slightly steepens the prediction, but within experimental error still remains consistent with our data, see Supplementary Material. 
Eq.\ (\ref{correlation}) also  explains  the observed steep increase of the average velocity $\left<U\right>$ above the threshold temperature (Fig.\ \ref{fig:fig2}c), as both  $\Delta T$ (Fig.\ \ref{fig:fig4}c) and $\theta$ (Fig.\ \ref{fig:fig3}) increase with the surface temperature $T_s$.

In summary, our experiments reveal a surprising phenomenon of self-propulsion that includes a volatile droplet and a uniformly heated wettable, conducting surface. Using  infra-red imaging we demonstrate that the spontaneous contraction of droplets coupled with spontaneous symmetry breaking of the Marangoni  flow inside the droplet results in  self-propulsion. Though results are mainly discussed in the context of ethanol on a sapphire substrate, we found that the self-propulsion is also observed for a variety of droplet-substrate pairs. These findings (exemplified in Table-\ref{tab:table}) 
also verify the condition  $k_\mathrm{R} \geq 4.7$ for the occurrence 
of the investigated phenomenon  
and  highlight its generic and robust nature, which can be a major hinderance for precise droplet deposition
required for many applications such as inkjet printing. The next step therefore must be a detailed parameter study of the autothermotaxis of volatile droplets.


\noindent
{\it Acknowledgements:} We thank Steffen Hardt, Olga Shishkina and Jiaming Zhang for fruitful discussion on the subject. NWO is kindly acknowledged for financial support.

\bibliographystyle{prsty_withtitle}
\bibliography{Bibliography}


\end{document}